\documentclass[letterpaper]{article} 
\usepackage{aaai2026}  
\usepackage{times}  
\usepackage{helvet}  
\usepackage{courier}  
\usepackage[hyphens]{url}  
\usepackage{graphicx} 
\usepackage{natbib}  
\usepackage{caption} 
\usepackage{amsmath}
\usepackage{amssymb}
\usepackage[utf8]{inputenc} 
\usepackage{booktabs}
\usepackage{longtable}
\usepackage{makecell}
\usepackage{tabularx}
\usepackage{enumitem}
\usepackage{multirow}

\DeclareUnicodeCharacter{2212}{\textminus}

\title{YTShorts-92K: An Interpretable Multimodal Dataset of YouTube Shorts}
\author{
    Celian Charrin\textsuperscript{\rm 1},
    Sidonie Abuzakuk\textsuperscript{\rm 2},
    Artúr Vlaszov\textsuperscript{\rm 3, 4},
    Nándor Birher\textsuperscript{\rm 3, 5},
    Roland Molontay\textsuperscript{\rm 3, 4}
}
\affiliations{
    \textsuperscript{\rm 1}Grenoble INP - ENSE3\\
    \textsuperscript{\rm 2}Grenoble INP - Phelma\\
    \textsuperscript{\rm 3}Department of Stochastics, Institute of Mathematics, Budapest University of Technology and Economics\\
    \textsuperscript{\rm 4}Institute of Biostatistics and Network Science, Semmelweis University\\
    \textsuperscript{\rm 5}Pázmány Péter Catholic University
}

\begin{document}

\maketitle

\begin{abstract}
Short-form video has become one of the dominant formats of online content, yet large-scale datasets enabling the analysis of what makes these videos successful remain limited. Existing resources often focus on post-publication popularity signals or rely on non-interpretable representations such as high-dimensional embeddings. Here, we introduce YTShorts-92K, a dataset of 92,654 YouTube Shorts collected from 345 creators and spanning 12 manually defined content categories. Each video is enriched with an interpretable multimodal feature set covering metadata, textual information, visual characteristics from frames and thumbnails, and cross-modal consistency measures. We describe the collection methodology, data cleaning process, and feature extraction pipeline, followed by descriptive analyses of the diversity of creators, topics, and extracted features. This dataset offers a reusable resource for studying short-form video content, online attention, and multimodal content analysis.
\end{abstract}


\section{Introduction}

Over the past decade, short-form video has become one of the dominant forms of online media, driven by platforms such as TikTok, Instagram Reels, and YouTube Shorts \cite{modern_day_high,dominance_tiktok,short_stat}. These videos, typically lasting no more than a few minutes, are consumed through infinite-scrolling, recommendation-driven interfaces designed to maximize user engagement \cite{modern_day_high}. Billions of short-form videos are viewed worldwide every day, making it increasingly important to understand which content characteristics contribute to their success \cite{dominance_tiktok,short_stat}.

Addressing this question at scale requires datasets that are both large enough to capture the diversity of online content and sufficiently rich to describe the content itself. Existing resources for large-scale video popularity research generally fall into two categories. The first consists primarily of metadata-based datasets containing information such as view counts, upload dates, and channel identifiers \cite{early_views_pattern,popularity_svm}, which provide limited insight into the intrinsic content characteristics underlying popularity. The second relies on embedding representations of titles, thumbnails, or transcripts \cite{cover_title_large,title_prediction_behavior,thumbnail_caption_prediction}, which, despite their predictive power, remain difficult to interpret. 

To address these limitations, we present YTShorts-92K, a large-scale dataset of approximately 92,000 YouTube Shorts enriched with an interpretable multimodal feature set spanning metadata, textual, visual, and cross-modal characteristics. The dataset provides broad temporal, topical, and creator diversity, enabling downstream research on video popularity prediction, content–popularity relationships, and the analysis of which interpretable content characteristics are associated with success. This article focuses on the dataset itself, detailing its collection pipeline and feature extraction methodology.
For social scientists, the semantic and emotional features can allow for the study of digital behavior at scale. For researchers in multimedia and/or marketing, the extracted thumbnail and frame features provide insights into visual hook optimization and creator branding. Furthermore, because the dataset is released in a structured, tabular format, it is suited for easy modeling tasks without the need to build complex data architecture beforehand.

\section{Related Work}

This section reviews existing studies and datasets for video content analysis, focusing on metadata-based resources, interpretable multimodal features, and embedding-based representations.

\paragraph{Metadata-based video resources}~\\
Many publicly available resources for studying video popularity rely almost exclusively on platform metadata, such as view counts, upload dates, channel identifiers, and, in some cases, post-publication view trajectories \cite{early_views_pattern,popularity_svm,YouNiverse}. These datasets are valuable for analyzing popularity dynamics but provide only a limited description of the video itself. When visual information is included, it is generally restricted to a small number of handcrafted thumbnail descriptors, such as colorfulness, contrast, or saturation \cite{popularity_svm}. As a result, they do not capture the rich combination of textual and visual characteristics that define the content presented to viewers before publication. To overcome the limitations of single-platform or single-indicator datasets, recent studies like the XS-Video dataset construct graphs in order to capture multi-dimensional interactions (e.g., shares, collects, comments) across multiple platforms \cite{xs_video_2026}. Similarly, datasets such as SnapUGC employ real user watch times to better model early-stage viewer retention~\cite{vquala2025}.

\paragraph{Interpretable multimodal content descriptions}~\\
Another category of studies has focused on extracting interpretable content features associated with user engagement. For long-form YouTube videos, studies have examined thumbnail properties (e.g., colorfulness, facial presence, facial expressions) together with textual characteristics such as title length, capitalization, emojis, or sentiment \cite{click_youtube,features_engagement}. For short-form video, existing research has focused on relatively small datasets, emphasizing factors such as opening hooks, trend alignment, and metadata optimization \cite{communicative_power,curious_shorts,trends_tiktok,virality_youtube}. However, these studies typically investigate only a subset of modalities in isolation and, to our knowledge, none releases a reusable dataset that jointly describes short-form videos through an interpretable multimodal feature set.

\paragraph{Embedding-based content representations}~\\
More recent work represents titles, thumbnails, and transcripts using transformer-based embeddings \cite{cover_title_large,title_prediction_behavior,thumbnail_caption_prediction,performance_marketing,abu2016youtube8m}. These representations capture rich information but are inherently difficult to interpret. Furthermore, the extracted embeddings and associated feature tables are generally not released as reusable research resources, limiting their value for studies seeking to analyze the contribution of specific content characteristics.

\begin{figure*}[pt]
    \centering
    \includegraphics[width=1\textwidth]{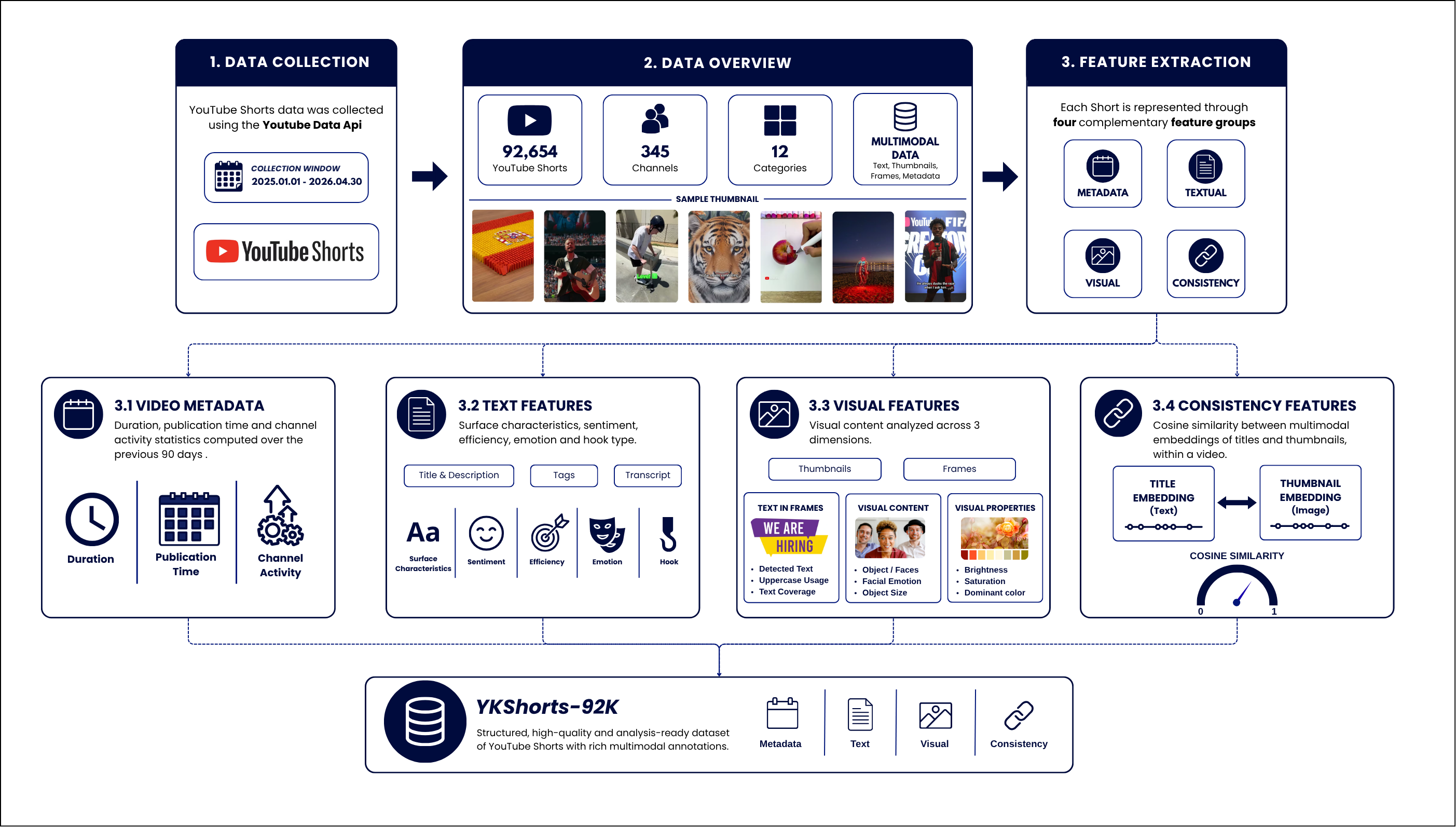}
    \caption{Complete framework for the creation of YTShorts-92K}
    \label{fig:framework_dataset}
\end{figure*}

\paragraph{Research gap}~\\
Table~\ref{tab:research_gap} summarizes how the dataset presented in this paper differs from existing resources. To our knowledge, no publicly described dataset simultaneously provides a large-scale collection of YouTube Shorts together with interpretable multimodal features extracted exclusively from pre-publication content. Figure~\ref{fig:framework_dataset} displays the complete pipeline used to create YTShorts-92K.

\begin{table*}[t]
\centering
\footnotesize
\setlength{\tabcolsep}{3pt} 
\caption{Comparison of the presented dataset with existing studies on video popularity.}
\label{tab:research_gap}
\begin{tabular}{llccccc}
\toprule
\textbf{Theme} & \textbf{Study} &
\makecell{\textbf{Short-}\\\textbf{form}} &
\makecell{\textbf{Interpretable}\\\textbf{features}} &
\textbf{Multimodal} &
\makecell{\textbf{Pre-}\\\textbf{publication}\\\textbf{only}} &
\makecell{\textbf{Large scale}\\\textbf{($>$50k)}}\\
\midrule

\multirow{5}{*}{\begin{tabular}{l} Metadata-based video \\ resources \\ \end{tabular}}
& \cite{early_views_pattern} & $\times$ & $\times$ & $\times$ & $\times$ & $\times$ \\
& \cite{popularity_svm} & $\times$ & \checkmark & \checkmark & $\times$ & $\times$ \\
& \cite{YouNiverse}    & $\times$ & \checkmark & $\times$  & $\times$ & \checkmark \\
& \cite{xs_video_2026}       & \checkmark & $\times$ & \checkmark & $\times$ & \checkmark \\
& \cite{vquala2025}          & \checkmark & $\times$ & \checkmark & $\times$ & \checkmark \\
\midrule

\multirow{6}{*}{\begin{tabular}{l} Interpretable multimodal \\ content descriptions\\ \end{tabular}}
& \cite{click_youtube}             & $\times$ & \checkmark & \checkmark & \checkmark & \checkmark \\
& \cite{features_engagement}       & $\times$ & \checkmark & \checkmark & \checkmark & $\times$ \\
& \cite{communicative_power} & \checkmark & $\times$ & $\times$ & \checkmark & $\times$ \\
& \cite{curious_shorts}      & \checkmark & $\times$ & $\times$ & \checkmark & $\times$ \\
& \cite{trends_tiktok}       & \checkmark & $\times$ & $\times$ & \checkmark & $\times$ \\
& \cite{virality_youtube}    & $\times$ & $\times$ & $\times$ & $\times$ & $\times$ \\
\midrule

\multirow{5}{*}{\begin{tabular}{l} Embedding-based content\\ representations\\ \end{tabular}}
& \cite{cover_title_large}            & $\times$ & $\times$ & \checkmark & \checkmark & $\times$ \\
& \cite{title_prediction_behavior} & $\times$ & \checkmark & $\times$   & \checkmark & $\times$ \\
& \cite{thumbnail_caption_prediction} & $\times$ & $\times$ & \checkmark & \checkmark & $\times$ \\
& \cite{performance_marketing}        & $\times$ & $\times$ & \checkmark & \checkmark & $\times$ \\
& \cite{abu2016youtube8m}  & $\times$ & \checkmark & $\times$ & $\times$ & \checkmark \\
\midrule

\textbf{YTShorts-92K} & & \checkmark & \checkmark & \checkmark & \checkmark & \checkmark \\
\bottomrule
\end{tabular}
\end{table*}
\section{Dataset Construction}
\label{part:dataset_construction}

This section describes the construction of the dataset, including the data collection procedure, the inclusion criteria, the cleaning process, and the resulting dataset.

\subsection{Dataset Collection}

The dataset is collected using the YouTube Data API v3, which provides access to video and channel metadata.

\paragraph{Collection window and stabilization}
The collected videos were published between January~1,~2025 and April~30,~2026, covering a period of approximately 16 months. The data were collected between June~1 and June~5,~2026, approximately one month after the latest publication date. According to the YouTube Shorts Product Lead, Todd Sherman, the recommendation algorithm initially tests each video on a subset of the audience (the ``seed'') before deciding how widely to distribute it \cite{lifespan_shorts}. Engagement is consequently concentrated in the first few days after publication \cite{engagement_through_time}. Given this behavior, the one-month delay between the latest publication date and data collection was intended to reduce recency bias in the collected view counts by allowing sufficient time for views to stabilize.

\paragraph{Inclusion criteria} Three criteria were applied during collection:

\begin{enumerate}
  \item \textbf{Western cultural scope.} 
  Channels were selected from Europe and North America, broadly representing Western cultural contexts and predominantly using the Latin writing system. This choice helps limit linguistic and cultural heterogeneity within the dataset.

\item \textbf{Channel diversity.} 
Channels were selected to cover a broad range of subscriber counts and content topics, reducing selection bias and improving the diversity of the dataset. 

The baseline channel set is extracted from YouTube creator rankings like SocialBlade \cite{socialblade}. We choose to only keep western channels from these rankings as per the first criteria.
An experiment is then conducted in which a member of our team is asked to use the Youtube Shorts platform as intended. During this trial, this person is prompted to take note of each unique channel they come across. As to avoid algorithmic personalization, the subject's YouTube account is rebooted every 30 minutes. The experiment takes place over 3 hours. The goal is to create a non-biased list of the channels that YouTube deems typical of the platform as to end up with a sample dataset as representative as possible.
Finally, a primitive descriptive analysis was conducted as to identify weak spots in the dataset (subscriber ranges or topics not represented). Targeted searches were then performed in order to fill these gaps (example : searching \textit{"Sports"} in the Shorts tab and selecting a channel that is between 50k-100k subscribers).

  \item \textbf{Shorts-only filtering.} Only public YouTube Shorts are retained. Videos longer than 180 seconds are discarded at collection time, following YouTube's duration definition of Shorts. Because the API does not expose a reliable field identifying Shorts specifically, a secondary verification step (described in Subsection \ref{part: dataset_cleaning}) removes non-Shorts videos that nonetheless satisfy the duration constraint.
\end{enumerate}

Following these criteria, a total of \textbf{100,034} YouTube videos from \textbf{348} unique channels were scraped from the Youtube Data Api V3.

\paragraph{Transcripts and topic labels} Automatically generated transcripts, which constitute the majority of transcripts on the platform, are not exposed by the YouTube Data API and were instead collected using the youtube-transcript-api. Transcript timestamps are also used to compute speech-rate features.

The API's \textit{categoryId} field was found in preliminary analysis to be too coarse for content-based analysis. Content topics are therefore manually assigned at the channel level and propagated to all videos of that channel. These 12 categories, along given with their respective distribution in the dataset are as follows :

\begin{itemize}[noitemsep]
    {\small
    \item[--] Challenges / Pranks / Family (5.0 \%)
    \item[--] Lifestyle / Vlog (9.1 \%)
    \item[--] Beauty / Fashion (6.6 \%)
    \item[--] Sports (12.4 \%)
    \item[--] Tech / Gaming (18.5 \%)
    \item[--] Animation (3.9 \%)
    \item[--] Food / Cooking (5.7 \%)
    \item[--] Fitness / Health / Medical (5.0 \%)
    \item[--] DIY / Arts \& Crafts (10.5 \%)
    \item[--] Science / Education (5.7 \%)
    \item[--] Animals (3.3 \%)
    }
\end{itemize}

Finally, the metadata collected for every video are listed in Table~\ref{tab:youtube_metadata}.

\begin{table}[h]
\centering
\caption{Metadata collected for each video.}
\label{tab:youtube_metadata}

\begin{tabularx}{\linewidth}{l l X}
\toprule

\textbf{Scraping} & \textbf{Field} & \textbf{type} \\

\midrule

\multirow{12}{*}{Youtube Data API}
& video\_id & \textit{string} \\
& title & \textit{string}\\
& publishedAt & \textit{datetime} \\
& country & \textit{categorical}\\
& description & \textit{string}\\
& thumbnail URL & \textit{string} \\
& video URL & \textit{string}\\
& tags & \textit{string list} \\
& views & \textit{integer}\\
& likes & \textit{integer} \\
& comments & \textit{integer}\\
& duration & \textit{integer}\\
\midrule

\multirow{2}{*}{youtube-transcript-api}
& transcript & \textit{string}\\
& wpm & \textit{float}\\
\midrule

\multirow{1}{*}{Manual allocation}
& topic & \textit{categorical}\\

\bottomrule

\end{tabularx}
\end{table}

\subsection{Data Cleaning}
\label{part: dataset_cleaning}

Videos satisfying the Shorts duration limit but not actually published as Shorts are removed by parsing each video page with \textit{BeautifulSoup} and discarding videos whose canonical URL is not a \texttt{/shorts/} endpoint. This cleaned set forms the baseline for all subsequent feature extraction. 

After applying the cleaning procedure described above to the 100,034 videos, the dataset contains 95,756 YouTube Shorts collected from 345 creators spanning 12 manually defined content categories. We can note that this number is different from the 92,654 that will constitute the final dataset. This is mainly due to videos that were no longer publicly accessible at the time of feature extraction, as well as samples removed because of missing feature values.

\section{Feature Extraction}
\label{part:feature_extraction}

To characterize YouTube Shorts from multiple perspectives, we extract a set of interpretable features from four complementary sources: metadata, textual information, video frames, and thumbnails. 

Table~\ref{tab:feature_set_description} summarizes the complete set of extracted features. The following subsections describe the extraction methodology adopted for each feature category.

\begin{table*}[t]
\centering
\caption{Summary of the extracted feature categories.}
\label{tab:feature_set_description}

\begin{tabularx}{\textwidth}{llX}
\toprule
\textbf{Category} & \textbf{Feature Group} & \textbf{Description} \\
\midrule

Metadata &
Video metadata &
Duration, publication time, and channel activity statistics computed over the previous 90 days. \\
\midrule

\multirow{3}{*}{Text}
& Title features
& Surface characteristics, sentiment, keyword efficiency, emotion, and hook type. \\

& Description, tags, and transcript
& Transcript length, speech rate, description length, and number of tags. \\
\midrule

\multirow{3}{*}{Frames}
& Low-level visual features
& Motion, frame differences, entropy, brightness, and contrast. \\

& High-level visual features
& Text detection and human presence estimated using deep learning models. \\
\midrule

\multirow{5}{*}{Thumbnail}
& Textual features
& Detected text, uppercase usage, and text coverage. \\

& Object and facial features
& Objects, faces, demographic attributes, facial emotion, and relative object size. \\

& Color features
& Brightness, saturation, dominant color, and visual complexity. \\
\midrule

Consistency& Video consistency & Cosine similarity between multimodal embeddings of titles and thumbnails, within a video.\\

\bottomrule

\end{tabularx}
\end{table*}

\subsection{Metadata-Based Feature Extraction}

Metadata features are derived from the information provided by the YouTube Data API (Table~\ref{tab:youtube_metadata}).

Temporal information is obtained from the \textit{publishedAt} field using Python's \textit{datetime} library. Publication timestamps are intentionally kept in UTC rather than converted to local time zones in order to learn about the international habits of Youtube users.

Furthermore, to characterize the creator's recent activity, additional statistics are computed over the previous 90 days.

The extracted metadata features are:

\begin{itemize}
    \item \textbf{Video duration (int):} duration of the video in seconds.
    \item \textbf{Temporal features (int):} publication hour, day of the week, and month.
    
    \item \textbf{Mean views (float):} average number of views over the previous 90 days.
    \item \textbf{View variance (float):} variance of view counts over the previous 90 days.
    
\end{itemize}

\subsection{Textual Feature Extraction}

Textual information is extracted from four complementary sources: video titles, descriptions, tags, and automatically generated transcripts. Since the title is the primary textual element visible to users before watching a video, it is analyzed in greater detail.

First, surface features describing the structure of the title are extracted. We then apply several natural language processing models to infer semantic information, including sentiment, emotion, and engagement-oriented hook categories.

The extracted title-based features are:

\begin{itemize}

\item \textbf{Surface features (int):} title length, number of words, capitalization patterns, punctuation usage, and the presence of emojis, hashtags, numerical values, and currency symbols.

\item \textbf{Sentiment (float):} sentiment score computed on the title using the \textit{VADER sentiment analysis} model, ranging from -1 (negative sentiment) to +1 (positive sentiment)~\cite{vadersentiment}.

\item \textbf{Efficiency (float):} ratio of extracted keywords to total words in the title, computed using the \textit{YAKE} keyword extraction algorithm. This feature measures how information-dense or concise a title is \cite{yake}.

\item \textbf{Emotion (categorical):} emotion classification obtained using a \textit{DistilRoBERTa}-based model, assigning each title to one of the seven Ekman emotions (joy, sadness, fear, anger, surprise, disgust, contempt) \cite{distillbert}.

\item \textbf{Hook type (categorical):} classification of titles into predefined engagement-driven patterns using a \textit{BART}-based model \cite{bart}. When the confidence level of the model goes under 0.5, the video is classified as "No hook visible". The possible regular categories are:
    \begin{itemize}[noitemsep]
    \item  "Revealing a secret or mystery"
    \item "An unexpected outcome or plot twist"
    \item "A relatable situation or struggle"
    \item "A wholesome or cute moment"
    \item "A hard challenge or high-stakes game"
    \item "A visually satisfying or odd process"
    \item "Viral hack or internet trend"
    \item "Expert commentary or legal breakdown"

    \end{itemize}
\end{itemize}

Additional statistics are extracted from descriptions, tags, and transcripts:

\begin{itemize}
    \item \textbf{Transcript length (int):} number of words in the transcript.
    \item \textbf{Speech rate (float):} words per minute (WPM).
     \item \textbf{Description presence (bool):} indicates whether the video description is non-empty.
    \item \textbf{Tag presence (bool):} indicates whether the video contains at least one tag.

\end{itemize}

Since descriptions, tags, and transcripts are not consistently available for all Shorts, we restrict the extracted features to those with reliable coverage across the dataset. This ensures that the released feature table remains complete and comparable across all videos.

\subsection{Visual Feature Extraction}

Visual features are extracted from two complementary sources: the video frames and the thumbnail.

\subsubsection{Frame Feature Extraction}

The extraction pipeline directly streams video content from YouTube's content delivery network with the help of yt-dlp \cite{yt-dlp}. Video containers are decoded frame by frame using PyAV, a Python interface built on top of FFmpeg~\cite{pyav}.

Only the first three seconds of each video are processed. This choice is motivated by YouTube Shorts' primary engagement metric, Viewed vs. Swiped Away (VVSA) \cite{VVSA,VVSA_youtube}. Frames are sampled at a fixed rate of 10 frames per second, resulting in 30 frames per video. To enable large-scale processing, feature extraction is parallelized across six worker processes.

Low-level visual descriptors are computed across all sampled frames:

\begin{itemize}
\item \textbf{Mean and Maximum Frame Difference (float)}: Visual variation between consecutive frames is quantified using the pixel-wise absolute brightness difference between frame $t$ and frame $t-1$. Then the mean and maximum are computed over this set in order to acquire the features. Higher values indicate faster scene transitions and greater visual dynamism.

\item \textbf{Mean and Maximum Optical Flow (float)}: Dense optical flow is computed between each pair of consecutive frames using the method introduced by Gunnar Farnebäck \cite{polynomial_expansion}, which estimates pixel displacement between two images based on polynomial approximations of local neighborhoods. The resulting displacement field provides a vector representation of local motion, capturing both its direction and magnitude. For each pair of consecutive frames, we compute the mean magnitude of the displacement vectors to quantify the average motion between frames. The mean and maximum of these values are then computed over the first 3 seconds of each video, capturing its overall and peak movement intensity, respectively.

\item \textbf{Mean and Standard Deviation of Entropy (float)}: Shannon entropy is calculated from the grayscale intensity distribution of each frame:

\begin{equation}
    H = -\sum_{k=0}^{255} p_k \log_2(p_k)
\end{equation}

where $p_k$ denotes the probability of observing intensity level $k$. Entropy serves as a measure of visual complexity, while its standard deviation reflects temporal variations in scene complexity. The mean and standard deviation of the frame-level entropy values are then computed to obtain the final features.

\item \textbf{Mean Brightness and Pixel Variance (float)}: The mean and variance of pixel intensities are computed for each frame to estimate brightness and contrast. These features capture the overall lighting conditions throughout the first 3 seconds of the video.

\end{itemize}

To complement these handcrafted descriptors, deep learning models are applied to a subset of ten uniformly sampled frames.

\begin{itemize}
\item \textbf{Text Presence and Text Ratio (bool \& float):} Optical Character Recognition is performed using EasyOCR \cite{easy_ocr}. We measure both the fraction of frames containing visible text and the average proportion of text pixels per frame.

\item \textbf{Human Presence and Density (bool \& float)}: A pre-trained YOLO \cite{yolo_ultralytics} object  detector identifies human figures within each frame. From these detections, we derive the person presence ratio and the mean number of detected persons per frame, allowing us to distinguish between individual creator content and group-oriented scenes.

\end{itemize}

\subsubsection{Thumbnail Feature Extraction}

Although Shorts are primarily consumed through an infinite-scrolling feed where videos start playing immediately, thumbnails remain important in search results, channel pages, recommendations, and external sharing. Consequently analyzing thumbnail characteristics provides additional information about how the content is visually presented to potential viewers.

Three categories of interpretable features are extracted from each thumbnail.

\paragraph{Textual Features}

EasyOCR \cite{easy_ocr} is used to extract textual information from thumbnails:

\begin{itemize}
\item \textbf{Text Ratio (float):} proportion of the thumbnail covered by detected text.
\item \textbf{Text Presence (bool):} indicates whether text is detected.

\item \textbf{Uppercase Usage (bool):} presence of fully capitalized words.
\end{itemize}

\paragraph{Object and Facial Features}

A pre-trained YOLO model \cite{yolo_ultralytics} detects objects appearing in the thumbnail. The extracted features describe the presence of persons, animals, food-related objects, and the relative size occupied by detected objects.

Whenever at least one person is detected, DeepFace \cite{boosted_lightface} performs facial analysis.

\begin{itemize}
\item \textbf{Face Statistics:} number of detected faces, estimated age, and face size measurements.
\item \textbf{Demographic and Emotional Attributes:} estimated gender, race, and dominant facial emotion.
\end{itemize}

\paragraph{Color Features}

The overall visual appearance of the thumbnail is characterized using the following descriptors:

\begin{itemize}
\item \textbf{Brightness and Saturation (float):} mean brightness and saturation computed in the HSV color space.
\item \textbf{Dominant Color (RGB):} dominant RGB color estimated using K-Means clustering ($K=3$).
\item \textbf{Visual Complexity (float):} Shannon entropy computed from the grayscale intensity distribution of the thumbnail.
\end{itemize}

\subsection{Consistency Feature}
Beyond individual textual and visual features, we can extract a consistency-based feature designed to capture the relationships between different content modalities. 

A title or thumbnail may be effective individually, but the coherence between these elements can also influence a video's attractiveness \cite{similarity_computing}. To quantify such relationships, we rely on multimodal embeddings generated with Qwen 3 VL \cite{qwen3vl}, which project heterogeneous data type (text and images) into a shared vector space. Similarities between embeddings are then computed using cosine similarity. The following feature is extracted: 

\textbf{Title-Thumbnail Consistency (float):} cosine similarity between the embedding of a video's title and the embedding of its thumbnail. This feature measures how closely the textual and visual content of a video are related.

\section{Descriptive Analysis}

After collecting the videos, merging all extracted features, and removing samples with missing values, the final dataset contains 92,654 YouTube Shorts with 80  extracted features.

In this section, we provide an overview of the dataset composition and report descriptive statistics to characterize its structure, diversity, and content distribution.

\subsection{Dataset Composition}

\begin{figure}[t]
    \centering
    \includegraphics[width=1\linewidth]{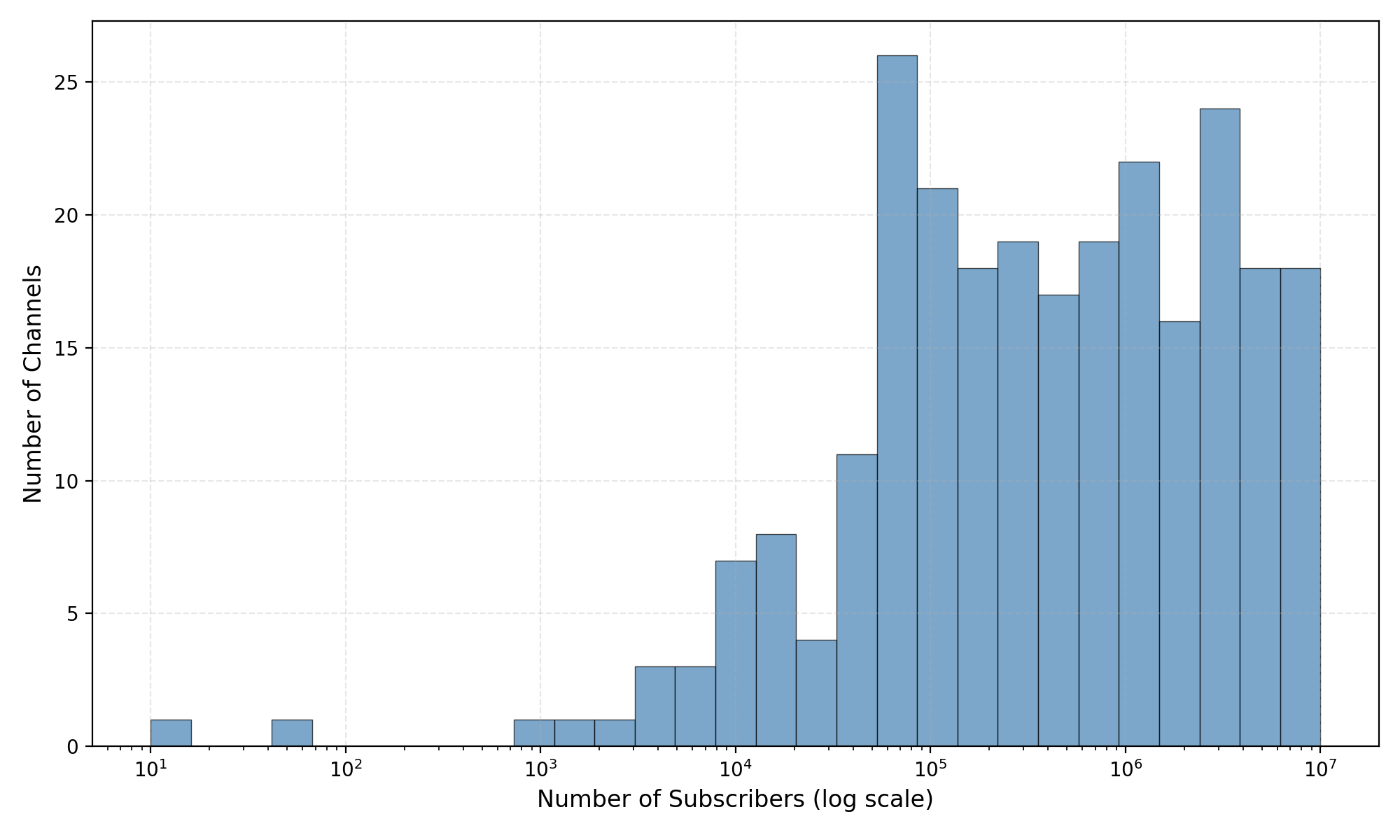}
    \caption{Distribution of the collected YouTube channels according to their number of subscribers (log scale).}
    \label{fig:channel_distribution}
\end{figure}

The dataset was intentionally designed to include creators covering a broad range of audience sizes. Figure~\ref{fig:channel_distribution} shows the distribution of channels according to their number of subscribers.

The collected channels range from only 11 subscribers to approximately 496 million subscribers, with a median of 1.26 million subscribers and an average of 9.57 million subscribers. Overall, 180 channels have at least one million subscribers, while the remaining 165 correspond to smaller creators.

Although the number of channels is nearly balanced between these two groups, larger channels naturally contribute more uploaded Shorts. Consequently, the final dataset contains 52,242 videos from channels with at least one million subscribers and 43,514 videos from smaller channels. This broad coverage reduces the risk that subsequent analyses are dominated by a particular creator profile and improves the representativeness of the dataset.

\subsection{Feature Correlations}

\begin{figure*}[pt]
\centering
\includegraphics[width=1\textwidth]{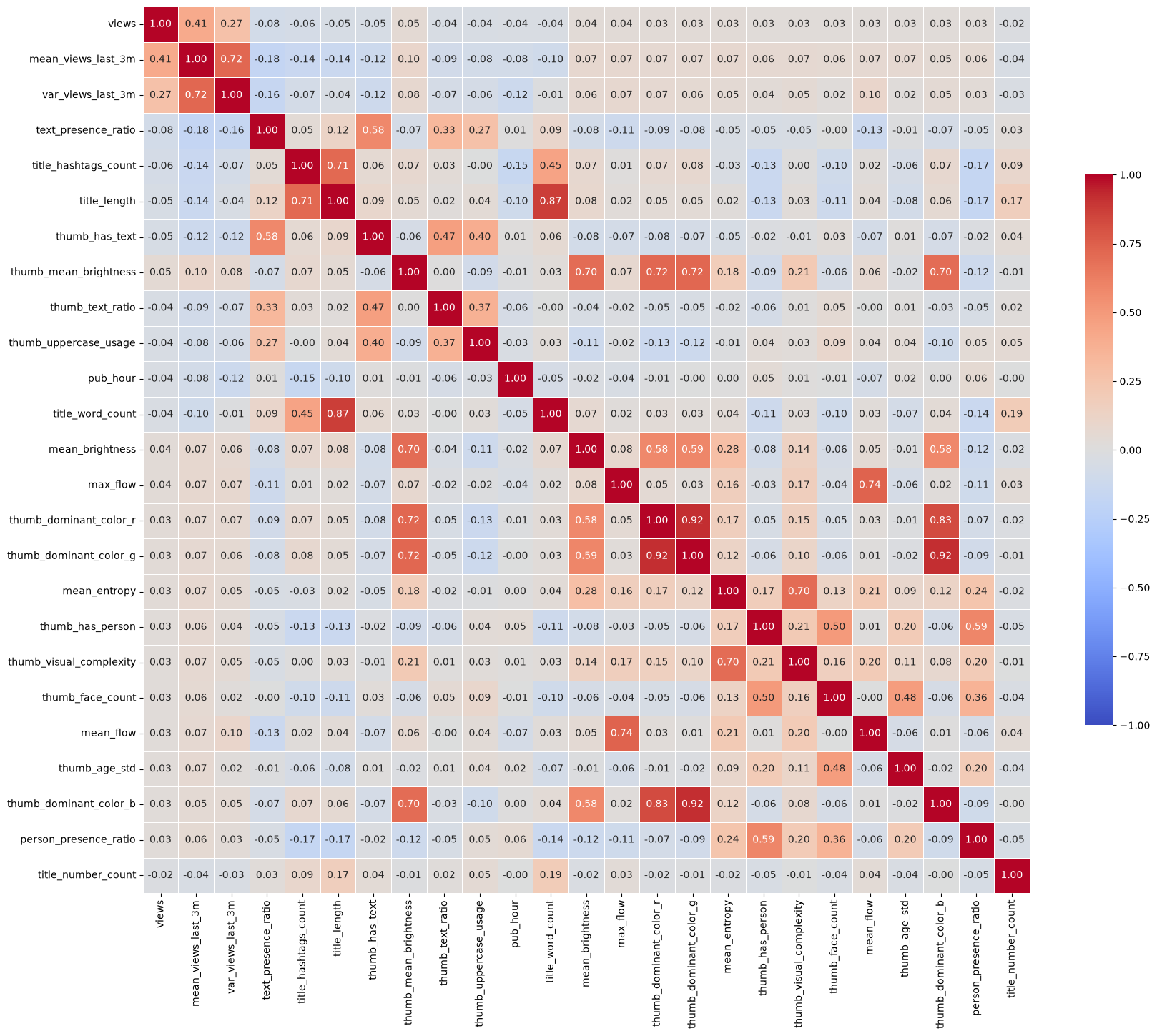}
\caption{Correlation matrix for a subset of extracted features, selected based on their correlations with \texttt{views}.}
\label{fig:correlation_matrix}
\end{figure*}

Overall, the extracted features exhibit relatively low pairwise correlations, suggesting that the different extraction pipelines capture complementary characteristics of YouTube Shorts. To keep the correlation matrix readable, Figure~\ref{fig:correlation_matrix} reports the subset of features showing the strongest correlations with the \texttt{views} variable.

As expected, the strongest correlations mainly occur between variables describing related underlying concepts. For example, the color descriptors \texttt{thumb\_dominant\_color\_r},\texttt{thumb\_dominant\_color\_g} and \texttt{thumb\_dominant\_color\_b} all seem well correlated, seeing values from ($r = 0.70$ to $r=0.92$). Similarly, \texttt{max\_flow} is strongly correlated with \texttt{mean\_flow} ($r=0.74$). Text-related variables show similar relationships: \texttt{length} is strongly correlated with \texttt{total\_words} ($r=0.87$) and with \texttt{hashtags} ($r=0.71$). 

We can observe however that, excluding the control features derived from it, the correlations between the features and the \texttt{views} variable are generally weak. Among the features shown in the Figure~\ref{fig:correlation_matrix}, visual features such as \texttt{text\_presence\_ratio} exhibits the strongest negative correlation ($r=-0.08$), followed by \texttt{hashtags} ($r=-0.06$), \texttt{length} ($r=-0.05$), and \texttt{has\_text} ($r=-0.05$). Overall, content characteristics display only limited linear associations with view count.

Some correlation between the engineered features are expected and do not necessarily indicate redundancy. Depending on the intended downstream application, users of linear models may choose to perform feature selection or dimensionality reduction to mitigate multicollinearity. In contrast, tree-based methods such as CatBoost, LightGBM, and XGBoost are generally robust to correlated predictors, allowing the complete feature set to be used without substantial performance degradation.

\section{Conclusion}

This paper introduced YTShorts-92K, a large-scale, interpretable multimodal dataset of 92,654 YouTube Shorts enriched with 80  features for studying short-form video content. By combining metadata, textual, visual, and cross-modal features, the dataset provides a detailed representation of videos beyond traditional popularity signals or non-interpretable embeddings.

Several limitations should be considered when using this resource. First, the dataset focuses on creators from Europe and North America, which may limit its generalizability to other cultural and linguistic contexts. Moreover, the creator-based collection strategy may introduce selection biases despite efforts to ensure diversity across topics and creator sizes.

Second, some potentially informative signals are unavailable. In particular, post-publication behavioral information, such as watch time, retention, or user interaction patterns, cannot be accessed through the public API and is therefore not included. Audio information is also unavailable, as the dataset does not include the audio tracks, despite music being a major component of YouTube Shorts. These missing signals may limit analyses aiming to fully characterize engagement dynamics and the factors contributing to video popularity.

Finally, the dataset reflects a specific period of platform evolution (January 2025–April 2026). Since content trends, recommendation systems, and audience preferences evolve over time, relationships identified using this resource may change as the short-form video ecosystem develops.

Future research can leverage this dataset for multimodal content analysis, creator strategy studies, and investigations of how visual and textual characteristics relate to online attention \cite{pulse_news}. Prior work has linked features such as thumbnail appearance, title structure, and visual-textual alignment to engagement on both long-form YouTube videos and short-form platforms \cite{click_youtube,features_engagement,thumbnail_caption_prediction,marketing_engagement,brand_engagement,similarity_computing}. Beyond content characteristics, creator persona and consistency have also been shown to influence audience response \cite{influencer_artistic_direction,parasocial_relationships}. While many of these studies focus on a limited set of content characteristics, the multimodal and interpretable feature representation released here enables visual, textual, and creator-related factors to be examined jointly and at scale. We could easily imagine different kind of new use for this data :

\begin{itemize}
\item Multimodal Consistency vs. Sensationalism: Does high semantic alignment between a video's title and thumbnail (title\_thumbnail\_consistency) result in higher median views compared to the use of aggressive engagement hooks (hook\_type) or high visual complexity (thumb\_visual\_complexity)?

\item Topic-Specific Attention Mechanisms: Do low-level visual dynamics (e.g., mean\_frame\_diff, mean\_flow) predict popularity differently across distinct content niches, such as fast-paced Gaming versus slower-paced Lifestyle content?

\item Emotional impact through demographics: How do specific emotional classifications in titles (title\_emotion) and facial expressions (thumb\_dominant\_emotion) interact with algorithmic reach across different creator sizes?

\end{itemize}


\section{FAIR Principles}
YTShorts-92k follows FAIR : \textbf{Findable, Accessible, Interoperable } and \textbf{ Reusable}.

\textbf{Findable}.The data and its corresponding metadata are hosted on Kaggle with a globally unique and persistent identifier :\\~
\url{https://www.kaggle.com/datasets/cliancharrin/ytshorts-92k}. The metadata is a ReadME file with all the necessary details to use the data and is searchable. Metadata are hosted at the same place as the data but still provide a link to them nonetheless.

\textbf{Accessible}. Metadata will always be accessible through the standardised and free http(s) protocol. All the URLs enable lawful access to YouTube and Google.

\textbf{Interoperable}. UTF-8 text, ISO-8601 UTC timestamps, numeric floats/ints; consistent emotion and topic vocabularies. JSON exports supported by common ML toolchains (e.g.,
Pandas, PyTorch, HuggingFace) and multimodal pipelines.

\textbf{Reusable}. Creative Common Zero licence. All data used is clearly identified and meets all community standards required. Possible to recreate end-to-end results with access to raw data.

\bibliography{reference}

\end{document}